# TOF-SIMS structural characterization of self-assembly monolayer of cytochrome b5 onto gold substrate

Satoka Aoyagi[a]*, Alain Rouleau[b], Wilfrid Boireau[b]

[a] Faculty of Life and Environmental Science, Shimane University, 1060 Matsue-shi, Shimane, 690-8504, Japan
[b] Institut FEMTO-ST, Université de Franche Comté, CNRS, 25044 Besançon, France
*aoyagi@life.shimane-u.ac.jp

Abstract

Orientation and three-dimensional structure of immobilized proteins on bio-devices are very important to assure their high performance. Time-of-flight secondary ion mass spectrometry (TOF-SIMS) is able to analyze upper surface of one layer of molecules. Orientation of immobilized proteins can be evaluated based on determination of a partial structure, representing ensemble of amino acids, on the surface part. In this study, a monolayer of cytochrome b5 was reconstituted onto gold substrate and investigated by surface plasmon resonance (SPR). After freeze-drying, the resulted protein self-assembly was evaluated using TOF-SIMS with the bismuth cluster ion source, and then TOF-SIMS spectra were analyzed to select peaks specific to cytochrome b5 and identify their chemical formula and ensembles of amino acids. The results from TOF-SIMS spectra analysis were compared to the amino acid sequence of the modified cytochrome b5 and three-dimensional structure of cytochrome b5 registered in the protein data bank. Finally, fragment-ion-generating parts of the immobilized-cytochrome b5 are determined based on the suggested residues and three-dimensional structure. These results suggest the actual structure and confirm the expected orientation of immobilized protein.

Keywords: cytochrome b5, protein immobilization, SAMs, three-dimensional structure of protein, TOF-SIMS, SPR

I. Introduction

There have been few methods to evaluate orientation of immobilized proteins, so far. Especially, it is very difficult to obtain direct information on immobilized-protein orientation because of its complex structure and random adsorption processes. Orientation and three-dimensional structure of immobilized proteins on bio-devices are crucial to assure their high performance [1]. In order to evaluate structural organisation, the surface structural characterization is necessary. Time-of-flight secondary ion mass spectrometry (TOF-SIMS) is able to analyze upper surface of one layer of molecules. The information depth for protein molecular fragment is considered to be approximately 1-2 nm under the static condition [2]. TOF-SIMS spectra contain peaks of fragment ions representing particular ensembles of amino acids on the surface part [3]. Therefore sophisticated analysis of TOF-SIMS spectra can determine the surface of the immobilized protein based on amino acid sequence.

In this study, cytochrome b5 self-assembly immobilized on the gold substrate was evaluated using TOF-SIMS and surface plasmon resonance (SPR). First step of the study is based on the reconstitution of a monolayer of proteins onto gold substrate. Protein of interest is a modified human cytochrome b5, having a unique cystein, i.e. a unique sulfhydryl residue at the periphery of the 3D structure. Bearing this sulfhydryl moiety, cytochrome b5 is able to (i) react with another cytochrome b5 to form dimeric species through disulfide bridge, (ii) crosslink with specific molecular spacers (linkers) and (iii) perform direct self assembly onto gold surface by chemisorption process leading to a packed and oriented self assembled monolayer (SAM) [4]. This last steep will occur at the surface of the sensor chip and will be analysed in real time by SPR measurements. Originality of the study is magnifying by on-chip complementary experiments onto the surface with TOF-SIMS. Finally these results are compared and orientation of the self-assembly monolayer of cytochrome b5 was proposed.

In order to select specific peaks of secondary ions related to each protein, mutual information defined by information theory [5-7] was applied to compare peaks among samples at once with considering intensity fluctuation depending on a TOF-SIMS measurement [8-10]. Intensities of a peak sometimes fluctuate by a TOF-SIMS measurement especially for bio-samples which are often composed of insulating materials. Analysis of spatial orientation of proteins is a challenge due to their complexes structures and conformations. Therefore differential comparison between a typical spectrum of a sample and that of a reference sample is sometimes inappropriate. According to values of the mutual information, important peaks are able to be selected comparing all TOF-SIMS spectra.

2. Experimental
*Sample Preparation*
Engineered cytochrome b5 was derived from human microsomal cytochrome b5 by genetic engineering resulting in the substitution of (i) the 26 C-terminal amino-acid residues by the –NGHHHH–COOH sequence and (ii) the serine 23 in Hb5(His)4 by cystein as previously described [11]. Cytochrome b5 was immobilized on the gold substrate in the following way. First step of the method is based on the reconstitution of a monolayer of proteins onto gold substrate. Protein of interest is a modified human cytochrome b5 which bears a unique cystein, i.e. a unique sulfhydryl residue. When cytochrome b5 is reduced by a reducing agent, it is able to react directly by chemisorption onto gold substrate and reconstitute a packed and oriented monolayer by direct self assembly. This steep will occur at the surface of the sensor chip and will be analysed by surface plasmon resonance measurements. Octylglucopyranoside (OG) and dithiothreitol (DTT) were purchased from by Sigma-Aldrich.

*SPR experiments*
BIAcore experiments were performed with Biacore™ 2000 apparatus. All of the experiments were carried out at 25 °C with a flow rate of 30 μl/min. Running buffer is composed of ultrapure water (Rathburn) and self assemblies of cytochrome b5 are processed in saline phosphate buffer (PBS, 100mM at pH7.4 with NaCl 50mM). Before injection, cytochrome b5 is reduced by DTT in excess (20/1 by moles) during ten minutes. Sensor chips used were bare gold substrates (SIA-kit provided by Biacore™). The immobilization degree of proteins and the level of interactions in Biacore technology are reported in a sensorgram (response unit (RU) versus time (s)). 1000 RU correspond to a shift in resonance angle of 0.1°. Calibration of the apparatus gives a correlation between the shift in angle and the mass deposition on the surface of the biochip of $0.1°–1$ ng/mm$^2$ [12]. The sensor chip is composed of four channels which have a surface of 1.4 mm$^2$. It is possible to perform SAMs with various density of protein in a window of 1 to 200 femtomoles/mm$^2$. All the process is performed in an aqueous environment. At the end of the assembly, pulses of detergent are performed to remove no-specific grafting.

*TOF-SIMS Analysis*
Positive ion spectra obtained with TOF-SIMS V (ION-TOF, Munster) using $Bi_3^+$ primary ion source, were acquired up to 1000 m/z while maintaining the primary ion dose less than $10^{12}$ ions/cm$^2$ to ensure static conditions [13]. The Raster size was 500 μm, and three times measurements were performed at different parts for each channel. All the spectra, composed of positive ion TOF-SIMS spectra, were calibrated to the $CH_3^+$, $C_2H_5^+$, and $C_3H_5^+$ peaks before data analysis. The mass resolution of $C_2H_3^+$ (m/z 27.02) is approximately 6000.

*Data Analysis*
Values of the mutual information [7-10] were calculated comparing the cytochrome b5-immobilized substrates with the substrates without protein. Peaks of secondary ions at m/Z = 40 to 500 were used for the calculation.

The results from TOF-SIMS spectra analysis were compared to the amino acid sequence of the modified cytochrome b5 and three-dimensional structure of cytochrome b5 registered in the protein data bank (http://www.rcsb.org/pdb/home/home.do).
Finally, fragment-ion-generating parts of the immobilized-cytochrome b5 are determined based on the suggested residues and three-dimensional structure.

## 3. Results and Discussion

The chip presents four channels including a reference channel (do not exposes to cytochrome b5 sample), and the 3 other channels have different amount of cytochrome b5 depending on the response of self assembly processes. The immobilization amount of cytochrome b5 on each channel was studied with SPR (Fig. 1). In order to improve self assembly of monomeric cytochrome b5 onto gold surface, the proteic sample is reduced by DTT just before injection. Thus, the sulfhydryl residue is really reactive and can more easily bind with gold substrate through chemisorption process. The different self assemblies reach a plateau at the end of the experiment, after 120 min of injection at 30 µl/min and give an average of 142 (+/- 15) femtomoles/mm$^2$ of immobilized cytochrome b5 after washing of the proteinchip with a non ionic surfactant, octyl glucopyranoside (OG). Very few amounts (less than 3% of coverage in triplicate experiments) of cytochrome b5 were absorbed by physisorption as demonstrated by this pulse. Table 1 shows coverage rate of each channel on the chip according to SPR results. Channel A does not contain cytochrome b5 but presents small amount of organic contaminations and/or drifting baseline after 180 min of experiments (145 RU i.e. less than 1 RU/min) and channel D adsorbs cytochrome b5 at maximum rate which is assumed to be 1 (around 2000 RU). At this stage, SPR experiments were stopped following a shut down procedure and an undocking of the chip without drying the channels. Thus the chip can be stocked in water prior to TOF-SIMS analyses.

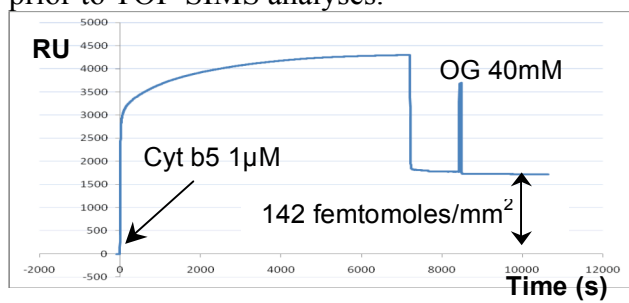

Figure 1 Sensorgramm of self assembly of reduced cytochrome b5 onto gold surface Channel B

Table 1 Cytochrome b5 coverage rate of each sample according to SPR

| Channel | Response Unit (RU) | Cyt. b5 Coverage (femtomoles.mm$^2$) | Coverage Rate (normalized to D) |
|---|---|---|---|
| A | 145 | / | 0 |
| B | 1730 | 142 | 0.89 |
| C | 1520 | 124 | 0.78 |
| D | 1930 | 158 | 1 |

According to TOF-SIMS spectra, differences among the channels are clearly shown as intensities of peaks related to the protein or the gold substrate. Comparing with all TOF-SIMS spectra several peaks of channels B, C and D are found to be greater than those of channel A.

Fig. 2 shows intensities of peaks increasing with cytochrome b5 and those of gold ion peaks. The intensities of gold ion peaks indicate the protein adsorption rate onto the substrate. On channel D the protein is immobilized at the highest density, and therefore the gold ion intensity is much lower than others. The other secondary ions of channels B, C and D in Fig. 2 are much higher than those of channel A, because these peaks of secondary ions are related to cytochrome b5. These secondary ions of channels B, C and D in Fig. 2 decrease with concentrations of cytochrome b5 samples.

Gold and silver often show somewhat enhancement effect of secondary ion intensity [14-16]. The enhancement effect caused by the gold substrate is the smallest in case of channel D because the substrate surface is covered with the protein at the highest rate, and therefore the intensities of protein-related peaks are less than those of channels B and C instead of the highest density of the immobilized protein. In addition, the intensities of protein-related peaks of channel B are greater than those of channel C, because the influence of coverage rate on channels B and C will be greater than the gold enhancement effect. In addition, according to the images of these secondary ions (data not shown), the gold ion image clearly shows that the gold ion distribution on the channel D area is less than others. The images of other secondary ions indicate that these fragment ions are much less generated from the channel A area.

Table 2 shows chemical formula of the fragment ions in Fig. 2. They are considered as specific fragment ions generated from cytochrome b5. The residues or residue combinations producing the fragment ions specific to cytochrome b5 shown in Table 2 suggest the secondary ion producing parts of the immobilized-cytochrome b5 molecule. These parts are considered to be on the surface or near the surface of the molecule. In order to consider three-dimensional structure of the protein, the protein having the same or similar amino acid sequence of the modified cytochrome b5 was found in the Protein Data Bank.

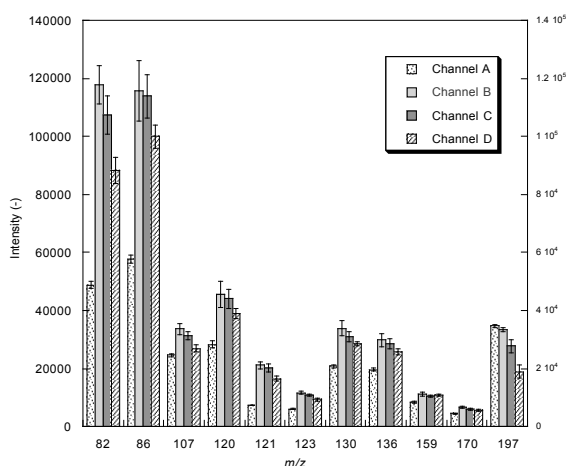

Figure 2 Peak intensity of each secondary ion (n = 3). Error bars show standard deviations.

Table 2 Chemical formula and possible residues of peaks

| m/z | Chemical formula | residues |
|---|---|---|
| 82.04 | C6H4N2 (82.053) | [R] |
| 86.09 | C5H12N (86.097) | [(L,I,K)] |
| 107.05 | C5H5N3 (107.048) | [(R,H)] |
|  | C4H11OS (107.053) | [M] |
| 120.08 | C8H10N (120.081) | [(F,Y)] |
| 121.04 | C5H5N4 (121.051) | [R] |
| 123.06 | C5H7N4 (123.067) | [R] |
| 130.06 | C9H8N (130.066) | [W] |
| 136.08 | C8H10NO (136.076) | [Y] |
|  | C5H14NOS (136.079) | [M] |
| 159.09 | C10H11N2 (159.092) | [W] |
| 170.06 | C11H8NO (170.06) | [W] |

The amino acid sequence of 1EHB.pdb (cytochrome b5) which has the similar sequence of the modified one is the following:
A(3)VKYYTLE(10)EIQKHNNSKS(20)TWLILHYKVY(30)DLTKFLEEHP(40)GGEEVLREQA(50)GGDATENFED(60)VGHSTDARE(69) LSKTFIIGELHPDDR (84 residues).
Underlined parts are different from the modified cytochrome b5, and numbers after residues are the number shown in the 1EHB figure. These numbers correspond to the following numbers for the residues in the modified cytochrome b5:

MAEQSDEA(3)VKYYTLEEIQKHNHCKS(20)TWLILHHKVY(30)DLTKFLEEHP(40)GGEEVLREQ(50)AGGDATENFED(60)VGHSTDARE(69)

The residues in Table 2 and their numbers in the sequence of 1EHB.pdb are the followings:
M, Y 6,Y 7,L 9, I 12, K 14, H 15, H 17, K 19, W 22, L 23, I 24, L 25, H 26, H27, K 28, Y 30, L 32, K 34, F 35, L 36, H39, L 46, R 47, F 58, H 63, D 66, R 68

Since the first methionine (M) is not included in the sequence of 1EHB, it does not have a number. The modified cytochrome b5 molecule is immobilized at cystein (C 23), C 23 is located at the same position of S 18 in 1EHB. According to the 3D structure (1EHB), residues Histidine (H 17), Lysin (K 19), Leucine (L 36), and Arginine (R 47) will not produce secondary ions because they are in internal parts and blocked by many other atoms. Even though the 3D structure might be different from the real one, residues H 17 and K 19 do not produce secondary ions under static-SIMS condition when the protein is immobilized at C 23.

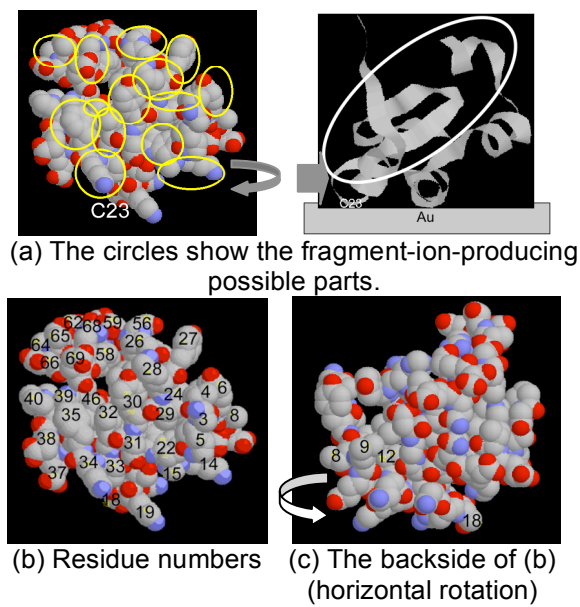

(a) The circles show the fragment-ion-producing possible parts.

(b) Residue numbers   (c) The backside of (b) (horizontal rotation)

Figure 3 Possible parts producing the fragment ions in table 2 (the modified "1EHB.pdb").

Considering these structural problems the fragment ion producing parts are determined as shown in Fig. 3. The side shown in the circled areas of Fig. 3 (a) contains the most of residues generating the fragment ions. Fig. 3(b) and (c) show the numbers of these residues in the sequence of 1EHB. Therefore it is indicated that this side would be the surface of the sample molecules. Moreover, this side should be the surface considering the immobilization processes of self-assembly monolayer of cytochrome b5. Further study is required to establish the evaluation method of such biomolecules with TOF-SIMS.

4. Conclusions

The self-assembly monolayer of cytochrome b5 on gold substrate was evaluated with SPR and TOF-SIMS. The protein monolayer densities at various conditions are clarified with SPR, and the surface structure of the protein molecule is determined based on the specific fragment ions from the protein in TOF-SIMS spectra. The orientation of the immobilized cytochrome b5 is indicated with this method and its three-dimensional structure is shown based on a registered structure on the Protein Data Bank. This method can be applied to various biological models including protein chip and opens the way to a more accurate investigation of biomolecular assemblies in nanobiotechnology and biosensors.

Acknowledgments
The authors appreciate Dr. Manabu Komatsu at Canon Inc for his useful and valuable help. This work was partly supported by Grant-in-Aid for Scientific Research (No. 19710096) from the Japanese Ministry for Education and Science and was supported by a fellowship PPF "Microtechniques for Proteomic" from Ministère de l'Enseignement Supérieur et de la Recherche (MESR).

References
[1] V. Mansuy-Schlick, R. Delage-Mourroux, M. Jouvenot, W. Boireau, Biosens. Bioelectron., **21** (2006) 1830–1837.
[2] R. Michel and D. G. Castner, Surf. Interface Anal., **38** (2006) 1386-1392.
[3] S. Aoyagi, M. Dohi, N. Kato, M. Kudo, S. Iida, M. Tozu, and N. Sanada, e-J. Surf. Sci. Nanotech., **4**, (2006) 614-618.
[4] W. Boireau, S. Bombard, M.A. Sari and D. Pompon, Biotech. Bioeng. **77** (2002), 225-231
[5] C.E. Shannon, W. Weaver, "The mathematical theory of information", University of Illinois Press, Urbana, IL, (1947).
[6] K.Eckschlager, V. Stepanek, K. Danzer, J. Chemometrics **4**, 195-216 (1990).
[7] T. Fujikura, K. Sakamoto, J. T. Shimozawa, Anal. Chim. Acta, **351,** 387 (1997).
[8] S. Aoyagi, M. Hayama. U. Hasegawa, K. Sakai, M. Tozu, T. Hoshi and M. Kudo, e-J. Surf. Sci. Nanotech., **1** 67-71 (2003).
[9] S. Aoyagi and M. Kudo, Biosens. Bioelectron., **20**(8) 1626-1630 (2005).
[10] S. Aoyagi, Y. Kawashima and M. Kudo, Nucl. Instr. Meth. Phys. Res. B, **232** 146-152 (2005).
[11] W. Boireau, A. Duncan, and D. Pompon, Meth. Mol. Biol. **300** (2005), 349-368.
[12] E. Stenberg, B. Persson, H. Roos, C. Urbaniczky, J. Colloid Interface Sci. **143** (1991), 513–526.
[13] G. Marletta, S. M. Catalano, S. Pignataro, Surf. Interface Anal., **16,** 407 (1990).
[14] L. Adriaensen, F. Vangaever, R. Gijbels, Anal. Chem. 76(22), (2004). 6777-6785
[15] A. Delcorte, N. Me´dard, P. Bertrand, Anal. Chem. **74**(19), (2002), 4955-4968
[16] A. F. M. Altelaar, I. Klinkert, K. Jalink, R. P. J. de Lange, R. A. H. Adan, R. M. A. Heeren, S. R. Piersma, Anal. Chem. **78**(3), (2006), 734-742